# Комплексная интеграция цифровых коллекций в информационное пространство научных исследований

И.А. Мбого, Д.Е. Прокудин, А.В. Чугунов

Университет ИТМО, Санкт-Петербургский государственный университет
irina.mbogo@gmail.com, hogben.young@gmail.com

## Аннотация

Настоящая статья рассматривает решение проблем аккумуляции и интеграции научных электронных коллекций в информационное пространство научных исследований. На основе анализа существующих стандартов и решений обосновывается выбор методологии и технологии представления электронных материалов научной конференции «Интернет и современное общество». Рассматривается концепция проекта интеграции создаваемой электронной коллекции в основные мировые и отечественные информационные системы и агрегаторы научной информации.

## 1. Введение

Развитие процессов информатизации научной деятельности в последнее время привело к расширению возможностей по оперативной публикации результатов актуальных научных исследований [8]. При этом основным средствами являются:

1. Различные программно-аппаратные издательские платформы для публикации в сети Интернет научных периодических сетевых изданий (в основном научных журналов). Это либо профессиональные решения крупных коммерческих издательств, либо различные «самописные» системы (например, научный электронный журнал «Аналитика культурологии» [6]), а также решения на базе свободно распространяемого программного обеспечения [1,4];
2. Сетевые электронные репозитории научной информации [2, 5].

В силу специфики в России в основном используются самостоятельные разработки или решения на основе свободно распространяемого ПО, так как основная масса изданий и репозиториев существует при поддержке различных научных сообществ и исходя из их возможностей как организационных, так и финансовых.

При этом формирование цифровых научных коллекций в значительной степени проходит спонтанно и зависит в первую очередь от необходимости обеспечить информационную поддержку исследовательской или образовательной деятельности. Это же в полной мере относится и к сборникам научных статей, а также материалам научных конференций. В основной своей массе статьи из этих источников не могут попасть даже такие базы научной информации как Научная электронная библиотека [10] или Киберленинка [9]. В основном в этом кроются такие причины как невозможность постфактум заключить авторские договора для соблюдения «чистоты» авторских прав; большого объёма сборников, метаданные которых необходимо вручную вводить в соответствующие системы (а в Научной электронной библиотеке это приходится делать сотрудникам издательств – а если сборник издавался организаторами конференции, то представителю университета или академической структуры).

К сожалению, в большинстве создаваемых электронных коллекций не только отсутствует механизм обмена метаданными, но и сами метаданные в каком-либо из международных форматов. Это, в свою очередь, делает проблематичным возможность автоматизации взаимодействия таких коллекций с основными репозиториями научной информации и создание условий для наилучшей индексации ресурсов в ведущих системах поиска научных публикаций Harzing's Publish or Perish, Google Scholar, Book Search, the ISI Web of Knowledge и других.

Поэтому для подавляющего числа отечественных научных сообществ актуальной задачей является решение проблемы формирования информационного пространства, адекватного современным требованиям, предъявляемым к инновационной научной деятельности.

Важный контекст для текущих изменений процесса научных исследований и разработок на базе интернет технологий задают международные инициативы, связанные с инициативой «Открытые архивы»:

- Будапештская инициатива «Открытый Доступ», 2001 г. (http://www.soros.org/openaccess/ru/read.shtml);
- Берлинская декларация об открытом доступе к научному и гуманитарному знанию, 2003 г. (перевод Э.М. Мирского - http://informika.ru/text/magaz/newpaper/messedu/2003/cour0311/200.htm);
- Международное соглашение «Берлин-3», 2005 г. (Berlin 3 Open Access: Progress in Implementing the Berlin Declaration on Open Access to Knowledge in the Sciences and Humanities. Feb 28th - Mar 1st, 2005, University of Southampton, UK – http://www.eprints.org/events/berlin3/outcomes.html);
- Международная петиция за гарантированный публичный доступ к результатам исследований, финансированных Европейской Комиссией, 2007 г. (http://www.ec-petition.eu/).

Важно отметить, что организации, подписавшие в 2003 г. Берлинскую декларацию об открытом доступе к научному и гуманитарному знанию, заявили о следующих намерениях:

a) стимулировать исследователей/получателей грантов публиковать свои работы согласно принципам парадигмы Open Archives Initiative;

b) стимулировать держателей культурного наследия поддерживать Open Archives Initiative, обеспечивая интеграцию интернет-ресурсов;

c) разрабатывать средства и способы оценки вкладов в Open Archives Initiative и сетевых журналов для того, чтобы поддерживать стандарты гарантии качества и лучшей научной практики;

d) добиваться, чтобы публикации в системе Open Archives Initiative признавались при присуждении ученых степеней и решении о занятии преподавательских должностей;

e) добиваться высокого качества вкладов в инфраструктуру Open Archives Initiative путем развития программных средств, поставки контента и создания метаданных или публикации индивидуальных статей.

В 2005 г. было принято дополнительное соглашение, названное «Берлин-3», призванное детализировать практические действия, которые рекомендуется выполнять исследовательским организациям в рамках Берлинской декларации. В этом соглашении обозначено, что в целях реализации Берлинской декларации научные организации (институты) должны требовать от своих научных сотрудников выкладывать в онлайновые архивы с открытым доступом электронные копии всех их опубликованных статей; и поощрять их публиковать статьи в журналах с открытым доступом к материалам, если подходящие журналы имеются (и обеспечивать сотрудникам поддержку в данных действиях).

В дальнейшем последовали публикации [13], еще более конкретизирующие сценарии поведения организаций и исследователей, действующих на базе Берлинской декларации. Научно-исследовательским организациям рекомендовалось для максимизации своего научного вклада взять на себя обязательство самоархивирования статей и материалов с результатами исследований. Организациям, финансирующим исследования (правительственные и частные), рекомендовалось сделать обязательным условием выделения грантов для финансируемых ими исследований требование самоархивирования результирующих материалов в открытых архивах организаций грантополучателей.

В России имеется ряд проектов, участники которых готовы к сотрудничеству в рамках настоящей инициативы.

Коллектив организаторов ежегодной научной конференции «Интернет и современное общество» (http://ims.ifmo.ru) поставил перед собой задачу решить эту проблему за счёт реализации проекта, идеология и применяемые технологические подходы которого лежат в русле международной инициативы создания открытых архивов научных публикаций (Open Archives Initiative - http://www.openarchives.org/). В состав этой международной системы входят так называемые Eprint Archive как тематические коллекции электронных документов, включающих научные публикации по разным тематикам. В рамках информационного пространства Open Archives имеется большое количество коллекций, представляющих гуманитарную науку. С этими коллекциями будет обеспечен взаимообмен в рамках настоящего проекта.

## 2. Решения для аккумуляции и распространения научного знания

На протяжении последних лет много говорилось об интеграции электронных библиотек на основе как библиотечных стандартов метаописаний, например MARK, RUSMARK и др., так и с использованием возможностей, предоставляемых информационными технологиями, например OAI-PMH. Следует отметить, что протокол OAI-PMH не всегда оправдывает возложенных на него задач по повышению уровня индексирования электронных библиотек ведущими агрегаторами научной информации. Тем не менее он является одним из основных механизмов интеграции метаданных, используемым достаточно большим числом агрегаторов.

Основными открытыми системами научного индексирования, например Google Scolar, просто выдвигается ряд требований к страницам описания публикации. Таким образом, можно выделить несколько актуальных задач: формирование корректных страниц публикаций внутри электронных библиотек, которые будут правильно индексироваться и восприниматься как публикация научными агрегаторами; большое количество полей для заполнения при пополнении электронных библиотек ведет к медленному пополнению коллекций. Актуальным является создание шаблонов для статей, их парсеров и инструментов импорта в электронные коллекции; также актуальной является задача загрузки материалов из электронной библиотеки в каталог РИНЦ на основании предложенных форматов.

Интересным, на наш взгляд, являются программные системы, которые сочетают в себе как функции репозиториев научных цифровых коллекций, так и возможности взаимодействия с внешними агрегаторами.

К таким системам, прежде всего, можно отнести Open Journal System (OJS), являющейся он-лайн издательской системой полного цикла, реализующей концепцию «электронного издательства» - компонента, обеспечивающего реализацию онлайновой инфраструктуры, предоставляющей возмож-ность гибких сценариев редакционной работы с научными текстами [12]. Основными особенностями этой системы, которые направлены на решение задач аккумуляции и распространения научного знания, являются:

- модуль «быстрой публикации», который позволяет создавать коллекции уже опубликованных научных статей. Через этот модуль вносится вся необходимая метаинформация: имена авторов, их аффиляция, названия и аннотации статей, списки литературы, ключевые слова и пр. к тому же информация может заполняться на нескольких языках в разных формах (например, на русском и английском языках);
- модуль импорта и экспорта статей и выпусков в формате XML, который поддерживает импорт для типа документа native.dtd. Поддерживаются корневые узлы <article>, <articles>, <issue> и <issues>, что позволяет интегрировать импортируемую метаинформацию (вместе с прикреплёнными файлами статей) в соответствующие агрегаторы;
- поддержка протокола OAI-PMH, что даёт возможность агрегировать информацию в различные информационные системы, аккумулирующие научную информацию.

Помимо этого, OJS обладает модулями экспорта метаданных в форматах:

- Erudit, определенном в виде DTD;
- CrossRef XML;
- PubMed XML для индексирования MEDLINE;
- XML для архивации в DOAJ.

Существуют, помимо рассмотренных, и другие возможности популяризации научного знания. Например, в последнее время в исследовательской среде набирают популярность так называемые «научные социальные сети» такие, как Academia.edu (http://academia.edu) или ReserchGate (http://reserchgate.org). Но, как правило, ответственность за наполнение этих информационных систем научными статьями, препринтами и результатами научных исследований лежит на пользователях этих сетей. Для издателей есть возможность зарегистрировать в системе свои научные периодические издания, а информация об их публикациях (как, например, в Academia.edu) попадает из новостных лент в формате RSS, представленных на сайтах изданий.

## 3. Использование общедоступных средств интеграции цифровых научных коллекций в мировое научное информационное пространство

В рамках реализации начатого проекта по интеграции материалов ежегодной научной конференции «Интернет и современное общество» создана проблемно-ориентированная научная электронная библиотека «Интернет и современное общество», аккумулирующая научные и научно-методические публикации по вопросам развития информационного общества, создания информационных ресурсов и систем для поддержки научных исследований в сфере гуманитарных наук. В качестве технологического решения была выбрана платформа OJS, на которой на сегодня размещены материалы конференций с 2011 по 2013 годы (http://ojs.ifmo.ru/index.php/IMS/) из весьма объемной информационной и методической базы, сформированной за 13 лет проведения всероссийской научно-методической конференции «Интернет и современное общество» (http://ims.ifmo.ru/). Тексты публикаций, размещенных в сборниках научных трудов (и сопутствующих изданиях), являются основой для создаваемого электронного репозитория. При этом полноценное использование возможностей «электронного издательства» будет осуществляться редакционной группой конференции при подготовке очередного сборника научных трудов и сопутствующих изданий.

Промежуточными результатами проекта можно считать следующие.

На основе возможности поддержки протокола OAI-PMH произведена автоматическая интеграция метаданных электронных публикаций (представляющих собой как электронные версии научных статей и тезисов конференции, так и препринтов, не имеющих бумажных аналогов) со «вторичными» научными информационно-поисковыми системами. Первой из них стала российская система «Соционет» (http://socionet.ru/), обеспечивающая информационную поддержку научно-образовательной деятельности во социогуманитарных и экономических научных дисциплинах. Эта система выполнена в рамках международных инициатив RePEc [10] и Open Archives Initiative [3] и представляет собой платформу для создания информационных ресурсов и сервисов, адресованных профессиональным научным сообществам. После создания в пространстве «Соционет» соответствующих профилей [7] была произведена привязка их к электронной коллекции материалов научной конференции «Интернет и современное общество» путём передачи администраторам системы ссылки на точку входа для индексирования электронного архива (http://ojs.ifmo.ru/index.php/index/oai). По протоколу OAI-PMH происходит автоматический сбор записей, содержащих метаданные статей.

При этом обновляются как существующие записи (если они были отредактированы при обнаружении ошибок ввода и неточностей), так добавляются новые записи (при добавлении в электронную коллекцию новых или архивных материалов). Эта коллекция была создана в системе «Соционет» 11 октября 2014 года и насчитывает на сегодня 160 записей (при статистике просмотров/загрузок 4/3).

Вслед за этим был сформирован запрос на индексацию электронной коллекции материалов конференции в Академию Google (http://scholar.google.ru/). Академия google рекомендует использовать последние версии программных продуктов Eprints (eprints.org), Open Journal System (OJS), DSpace (dspace.org). В остальных случаях необходимо выполнять технические требования Google. Академия Google не использует для индексации сайтов протокол OAI-PMH в полном объёме, поэтому используется обычный механизм обхода сайта для индексации страниц и построения карты сайта, хотя имеется дополнительный механизм, улучшающий индексацию, основанный на анализе метаданных, что и реализовано при регистрации типов ресурсов.

В случае с электронным репозиторием материалов конференции «Интернет и современное общество» на соответствующей странице был выбран тип сайта «Open Journal System (OJS)» (https://support.google.com/ scholar/troubleshooter/2898950#ts=2976969) и далее введён web-адрес репозитория (http://ojs.ifmo.ru). На момент подготовки настоящей статьи Академия Google не проиндексировала электронную коллекцию, так как на это (как указано на её сайте) требуется от 4 до 6 недель.

Также материалы конференции «Интернет и современное общество» за 2013 год были успешно размещены в Научной электронной библиотеке (http://elibrary.ru/item.asp?id=21718151). При этом пришлось вручную вводить все метаданные, используя инструмент XML-разметки «Артикулус» (доступен только по внутренней ссылке из авторизованного пространства издателей, заключивших соответствующий договор с НЭБ). Особенностью представления информации в НЭБ состоит в том, что нет необходимости размещать в этой системе полные тексты статей, поэтому были размещены ссылки на статьи, расположенные в электронной коллекции на сайте материалов самой конференции. Это позволило, с одной стороны, избежать дублирования информации (создание информационного «шума»), с другой – таким образом популяризировать сам созданный ресурс (чтобы для чтения текстов посещали саму коллекцию, а не сторонние по отношению к ней ресурсы).

## 4. Автоматмизация процессов обмена библиографической информацией между различными информационными системами

Одним из направлений реализации начатого проекта является разработка механизмов автоматизации процесса публикации научных статей в действующей системе представления научного знания, созданной на основе программного обеспечения с открытым кодом OJS (http://ojs.ifmo.ru/ims/) с дальнейшей автоматизацией процессов размещения метаинформации в ведущих агрегаторах научной информации. Это комплексная задача, включающая два этапа.

**Первый этап**

Разработка шаблона представления материалов (статей и тезисов докладов) на ежегодную Всероссийскую конференцию «Интернет и современное общество» для тестовых редакторов, который по-зволит сохранять метаинформацию о статье/тезисах в XML-файл в формате для импорта в системе Open Journal System. При этом должно учитываться корректное формирование описания публикации, соответствующее мета-тэгам XML-формата. Например, ниже приведён фрагмент файла статьи в XML-формате, который формируется модулем экспорта системы OJS:

```
<article><title locale="ru_RU"> Развитие ком-
плексных информационных систем поддержки
междисциплинарного научного направления:
решения и разработки </title> {Аннотация}
<indexing><subject locale="ru_RU">
информационно-коммуникационные техноло-
гии; информатизация научной деятельности; ин-
формационные системы
</subject></indexing><author pri-
mary_contact="true"><firstname>Ирина</firstnam
e><middlename>Анатольевна</middlename><last
name>Мбого</lastname><country>RU</country><
email>irina.mbogo@gmail.com</email><biography
locale="ru_RU">зав. лабораторией Санкт-
Петербургского государственного университета,
ведущий программист Центра технологий элек-
тронного правительства Санкт-Петербургского
национального исследовательского университета
информационных технологий, механики и опти-
ки.</biography></author><author><firstname>Дми
трий</firstname><middlename>Евгеньевич</midd
lename><lastname>Прокудин</lastname><country
>RU</country><email>hogben@philosophy.pu.ru<
/email><biography locale="ru_RU">докт. филос.
наук, доцент кафедры логики Санкт-
Петербургского государственного университе-
та.</biography></author><pages>178-
183</pages><galley
locale="ru_RU"><label>PDF</label><file><embed
filename="DL03Mbogo.pdf" encoding="base64"
mime_type="application/pdf">{Код файла в фор-
мате pdf}
</embed></file></galley></article>
```

Исходя из этого необходимо в шаблоне определить поля, которые при сохранении будут попадать в соответствующие мета-тэги:

```
<article><title locale="ru_RU"> - название статьи
на русском языке;
 - аннотация статьи на
русском языке;
<indexing><subject locale="ru_RU"> - ключевые
слова на русском языке;
<author primary_contact="true"> <firstname> - имя
первого соавтора;
<lastname> - фамилия первого соавтора и т.д.
```

**Второй этап**

Создание парсера для формирование XML-файлов на основе DTD или schema, предлагаемого РИНЦ для автоматического размещения информа-

Санкт-Петербург, 19 - 20 ноября 2014 г.                                                                 51

ции сборников статей научной конференции в Научной электронной библиотеке (http://elibrary.ru). Для этого необходимо произвести разбор форматов XML-файлов статей и выпусков модуля импорта/экспорта Open Journal System и форматов XML-файлов проектов системы Articulus (Научная электронная библиотека). После этого будет произведено построение набора правил конвертирования форматов XML-файлов между Open Journal System и Научной электронной библиотекой.

Ниже прилагается фрагмент кода XML-файла, формируемого в системе разметки Articulus:

```
<article>
    <pages>178-183</pages>
    <artType>PRC</artType>
    <authors>
        <author num="001">
            <individInfo lang="RUS">
                <surname>Мбого</surname>
                <initials>И.А.</initials>
                <orgName>Санкт-Петербургский государственный университет</orgName>
            </individInfo>
        </author>
        <author num="002">
            <individInfo lang="RUS">
                <surname>Прокудин</surname>
                <initials>Д.Е.</initials>
                <orgName>Санкт-Петербургский государственный университет</orgName>
            </individInfo>
        </author>
    </authors>
    <artTitles>
        <artTitle lang="RUS">Развитие комплексных информационных систем поддержки междисциплинарного научного направления: решения и разработки</artTitle>
        <artTitle lang="ENG">Evolution of Complex Information Systems for the Interdisciplinary Scientific Direction Support: Decisions and Development</artTitle>
    </artTitles>
    <abstracts>
        Обзорная статья представляет результаты исследования, которые были проведены с целью выбора базовых компонентов для создания комплексной модульной информационной системы для обслуживания междисциплинарного научного направления, ориентированного на изучение различных аспектов такого явления как «Информационное общество».
        The review represents results of research which were conducted for the purpose of a choice of basic components for creation of complex modular information system for service of the interdisciplinary scientific direction focused on studying of various aspects of such phenomenon as "Information society". At selection of components the interdisciplinary orientation of subject domain and the questions of construction of the open, available information system connected with it, allowing to carry out collective interaction between researchers were taken into account.
    </abstracts>
    <text lang="RUS"></text>
    <references>
        <reference>Логико-философские штудии [Электронный ресурс]. URL: http://ojs.philosophy.spbu.ru/index.php/lphs (дата обращения: 23.07.2013).</reference>
        <reference>Научная электронная библиотека [Электронный ресурс]. URL: http://elibrary.ru (дата обращения: 20.07.2013).</reference>
        <reference>Универсальная издательская платформа RAE Editorial System [Электронный ресурс] // URL: http://www.esrae.ru (дата обращения: 23.07.2013).</reference>
        <reference>Харнад С. Максимизация научного эффекта через институциональные и национальные обязательства самоархивирования для открытого доступа // Интернет и современное общество: труды IX Всероссийской объединенной конференции IMS-2006, СПбГУ, 14 - 16 ноября 2006 г. СПб., 2006. С. 22 - 29. URL: http://conf.infosoc.ru/2006/thes/harnad.pdf (дата обращения: 25.07.2013)</reference>
```

Как можно убедиться, существует соответствие мета-тэгов, которое позволит составить однозначные правила конвертирования. Например:

```
<artTitles><artTitle lang="RUS"> (НЭБ) соответствует <article><title locale="ru_RU"> (OJS).
```

## 5. Заключение

Поддержка основных стандартных протоколов обмена информацией (XML, OAI-MPH) системой OJS позволит зарегистрировать ресурсы, созданные в рамках предлагаемого проекта, в глобальных системах-сборщиках данных OAI (harvesters), например, таких как: scirus (http://www.scirus.com), myOAI (http://www.myoai.com), OAIster (http://oaister.umdl.umich.edu). Эти системы обеспечивают поиск научных публикаций, аккумуляцию и взаимообмен данными между большей частью университетов и исследовательских организаций всего мира.

Важным результатом проекта будет постепенное снижение оторванности российских исследователей, работающих в социально-гуманитарных отраслях науки, от мирового научного дискурса и информационного пространства. Использования методов и технологий интеграции ресурсов и сервисов, которые развиваются в рамках инициативы открытых архивов (OAI) позволяет обеспечить информационную интеграцию как с мировыми ресурсами (для тех публикаций, которые имеют набор метаинформации на английском языке: библиографическое описание, реферат или аннотацию, набор ключевых слов), так и с российскими научными электронными коллекциями, поддерживающими шлюзы на базе протоколов OAI. Решение данной задачи не только обеспечит постепенное появление в мировом научном информационном пространстве описаний российских научных публикаций по данной тематике (со ссылками на их полные тексты), но и распространит в России принципы современной культуры электронных научных публикаций, ставших стандартами де-факто в мировом научном коммуникационном пространстве.

## Литература

## Complex Integration of Digital Collections into Scientific Information Space

Irina A. Mbogo, Dmitry E. Prokudin,
Andrei V. Chugunov

The article considers the solution of problems of accumulation and integration of scientific electronic collections into information space of scientific researches. On the basis of the analysis of the existing standards and solutions the choice of methodology and technologies of representation of electronic materials of the scientific conference "Internet and Modern Society" locates. The concept of the project of integration of the created electronic collection into the main world and domestic information systems and aggregators of scientific information is considered.